\documentclass[5p,twocolumn,times]{elsarticle}
\journal{Physics Letters B}

\usepackage{graphicx}
\usepackage{amsmath}
\usepackage{amssymb}
\biboptions{sort&compress}

\DeclareRobustCommand{\<}[1]{\hspace{-0.11111em}#1\hspace{-0.11111em}}
\DeclareRobustCommand{\rtrim}[1]{#1\hspace{-0.11111em}}

\usepackage{liealg}

\DeclareRobustCommand{\grpsochain}{\grpso{5}\<\supset\grpso{3}}
\DeclareRobustCommand{\grpsutimes}{\grpsu{1,1}\times\grpso{5}}
\DeclareRobustCommand{\grpy}[2][]{\grpexpr{Y}{#1}{#2}}

\DeclareRobustCommand{\Lambdahat}{\hat{\Lambda}}
\DeclareRobustCommand{\calM}{\mathcal{M}}

\begin{document}


\title{
Phonon and multi-phonon excitations in rotational nuclei 
by exact diagonalization of the Bohr Hamiltonian
}

\author{M. A. Caprio}
\address{Department of Physics, University of Notre Dame,
Notre Dame, Indiana 46556-5670, USA}

\begin{abstract}
Exact numerical diagonalization of the Bohr Hamiltonian by
$\grpsutimes$ methods is used to obtain detailed quantitative
predictions for single-phonon and multi-phonon
excitations in well-deformed rotor nuclei.  Dynamical $\gamma$
deformation is found to significantly influence the predictions
through
its coupling to the rotational motion.  Basic signatures for the onset
of rigid triaxial deformation are obtained.  
\end{abstract}

\makeatletter
\def\keyword{%
  \def\sep{\unskip, }%
 \def\MSC{\@ifnextchar[{\@MSC}{\@MSC[2000]}}
  \def\@MSC[##1]{\par\leavevmode\hbox {\it ##1~MSC:\space}}%
  \def\PACS{\par\leavevmode\hbox {\it PACS:\space}}%
  \def\JEL{\par\leavevmode\hbox {\it JEL:\space}}%
  \global\setbox\keybox=\vbox\bgroup\hsize=\textwidth
  \normalsize\normalfont\def\baselinestretch{1}
  \parskip\z@
  \raggedright                         
  \ignorespaces}
\makeatother  


\maketitle


The Bohr collective Hamiltonian has served as a conceptual benchmark
for the interpretation of quadrupole collective dynamics in nuclei for
several decades~\cite{bohr1952:vibcoupling,bohr1998:v2}.  A tractable
scheme for numerical diagonalization of the Bohr Hamiltonian, the
algebraic collective model
(ACM)~\cite{rowe2004:tractable-collective,rowe2004:spherical-harmonics,rowe2005:algebraic-collective,rowe2005:radial-me-su11,caprioXXXX:gammaharmonic},
has recently been proposed, based on $\grpsutimes$ algebraic methods.
The need for such an approach arises since the conventional approach
to numerical diagonalization of the Bohr Hamiltonian, in a
five-dimensional oscillator
basis~\cite{gneuss1969:gcm,hess1980:gcm-details-238u,eisenberg1987:v1},
is slowly convergent and requires a large number of basis states to
describe a general deformed rotor-vibrator nucleus.  Consequently, it
has been necessary to apply varying degrees of approximation in
addressing the dynamics of transitional and deformed nuclei, as in the
classic rotation-vibration model~\cite{faessler1965:rvm} and rigid
triaxial rotor~\cite{davydov1958:arm-intro} treatments of the Bohr
Hamiltonian, or in more recent studies of critical
phenomena~\cite{iachello2001:x5,iachello2003:y5,bonatsos2007:gamma-separable-x5,bonatsos2007:gamma-separable-davidson}.

The ACM scheme, in conjunction with recent
progress in construction of the relevant $\grpsochain$
Clebsch-Gordan coefficients~\cite{caprioXXXX:gammaharmonic}, now
permits the diagonalization of the Bohr Hamiltonian for potentials of
essentially arbitrary stiffness, as considered in this letter.  The
Bohr Hamiltonian can consequently be applied, without approximation, to the full range of nuclear quadrupole
rotational-vibrational structure, from spherical oscillator to axial
rotor to triaxial rotor.
Specifically, the direct product basis obtained from an optimally chosen set of
$\mathrm{SU}(1,1)$ $\beta$ wave functions~\cite{rowe1998:davidson} and
the $\grpsochain$ spherical harmonics $\Psi_{v\alpha L
M}(\gamma,\Omega)$~\cite{rowe2004:spherical-harmonics} provides an
exceedingly efficient basis for numerical solution of the Bohr
Hamiltonian~\cite{rowe2005:algebraic-collective}.
For application to transitional and deformed nuclei, the method yields
order-of-magnitude reductions in the basis size needed for
convergence, as compared to diagonalization in a five-dimensional
oscillator basis.  The $\grpsutimes$ algebraic structure of the basis
facilitates construction of matrix elements for a wide variety of
potential and kinetic energy operators.

In this letter, detailed quantitative predictions for single-phonon
and multi-phonon excitations in deformed rotor nuclei are established
by exact numerical diagonalization of the Bohr Hamiltonian, making use
of newly-calculated $\grpsochain$ Clebsch-Gordan
coefficients~\cite{caprioXXXX:gammaharmonic}.  In the past,
interpretation of rotational phonon states within the Bohr description
has largely been at a schematic level (\textit{e.g.},
Refs.~\cite{gunther1967:166er-168er-beta,riedinger1969:152sm154gd-beta,warner1981:168er-ibm,wu1996:gamma-fragmentation-os,haertlein1998:168er-coulex}):
adiabatic separation of the rotational and vibrational degrees of
freedom is assumed, the $\beta$ and $\gamma$ excitations are taken to
be harmonic, and phonon selection rules are assumed for electric
quadrupole transitions.  These predictions are then adjusted by
spin-dependent band mixing~\cite{mikhailov1966:mixing-APS} with
\textit{ad hoc} mixing parameters.  Here, instead, we
explore the actual predictions of the Bohr Hamiltonian.  The
signatures for the onset of rigid triaxial deformation within the Bohr
framework are also considered.  Preliminary results were presented in
Ref.~\cite{caprio2008:acm-cgs13}.
\begin{figure*}
\begin{center}
\includegraphics*[width=0.66\hsize]{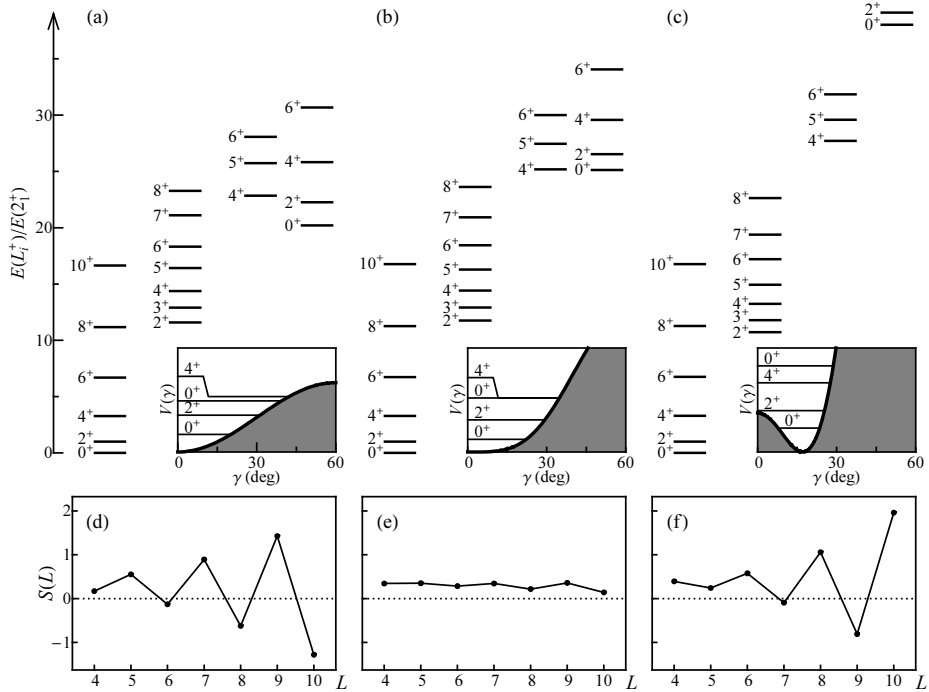}
\end{center}
\caption{Level schemes for the angular problem Hamiltonian~(\ref{eqn-gammarotor}), for
(a)~$\xi\<=0$ with $\chi\<=50$, (b)~$\xi\<=0.5$ with $\chi\<=100$, and
(c)~$\xi\<=0.8$ with $\chi\<=500$.  The potential is shown in the
inset, with the ground, quasi-$\gamma$, and quasi-$\gamma\gamma$ band
head energies indicated.  (d,e,f)~Staggering of level energies within the
quasi-$\gamma$ band, for the same three calculations, as measured by
the energy second difference $S(L)$.}
\label{fig-gammarotor-schemes}
\end{figure*}

The Bohr Hamiltonian~\cite{bohr1998:v2} is given, in terms of the
quadrupole deformation variables $\beta$ and $\gamma$ and Euler angles
$\Omega$, by
\begin{equation}
\label{eqn-bohr}
H=-\frac{\hbar^2}{2B}\biggl[
\frac{1}{\beta^4}\frac{\partial}{\partial\beta}\beta^4\frac{\partial}{\partial\beta}
-\frac{\Lambdahat^2}{\beta^2}\biggr]+V(\beta,\gamma),
\end{equation}
where $\Lambdahat^2$ is the angular $(\gamma,\Omega)$ part of the
Laplacian in five dimensions.  The essential aspect of the present ACM
solutions is that the angular degrees of freedom are treated in full,
including dynamical $\gamma$ deformation and its coupling to the
rotational motion.  In the context of small-oscillation approximations
for $\gamma$ (\textit{e.g.}, Ref.~\cite{bohr1998:v2,iachello2001:x5}),
the $\gamma$ dependence of the potential is simply taken as
$\rtrim\propto\gamma^2$, but for solution of the full problem the
$\gamma$ dependence must be defined more completely.  From the
symmetry properties of a quadrupole-deformed
nucleus~\cite{eisenberg1987:v1}, the potential energy
$V(\beta,\gamma)$ must be periodic in $\gamma$ (with period
$120^\circ$), and it must be symmetric about $\gamma\<=0^\circ$ and
$\gamma\<=60^\circ$.  The simplest potential of this form, with
minimum at $\gamma\<=0^\circ$, is
$V(\gamma)\<\propto(1-\cos3\gamma)$~\cite{rowe2004:tractable-collective},
as shown in Fig.~\ref{fig-gammarotor-schemes}(a,inset).

Let us first consider a problem
proposed by Iachello~\cite{iachello2003:y5}, with Hamiltonian
\begin{equation}
\label{eqn-gammarotor}
H=\Lambdahat^2 + \chi \bigl[ (1-\cos 3\gamma) + \xi \cos^2 3\gamma\bigr].
\end{equation}
Only the angular variables $(\gamma,\Omega)$ are considered, with
$\beta$ held fixed~\cite{fn-separation}.  The possible forms of the potential
$V(\gamma)$ and the results of illustrative calculations are
shown in Fig.~\ref{fig-gammarotor-schemes}.
For $\xi\<=0$, a simple
$(1-\cos 3\gamma)$ potential is obtained
[Fig.~\ref{fig-gammarotor-schemes}(a)], approximately harmonic
(locally $\rtrim\propto\gamma^2$) around
$\gamma\<=0^\circ$.  For $\xi\<=1/2$, the potential is more softly
confining in $\gamma$, with a quartic minimum ($\rtrim \propto
\gamma^4$) at $\gamma\<=0^\circ$
[Fig.~\ref{fig-gammarotor-schemes}(b)].  This case is termed
``critical'' in Ref.~\cite{iachello2003:y5}.  For $\xi\<>1/2$, the
potential has a minimum at some nonzero value of $\gamma$, given by
$\cos3\gamma_0\<=1/(2\xi)$ [Fig.~\ref{fig-gammarotor-schemes}(c)].
The basis functions for the diagonalization consist simply of the
$\grpsochain$ spherical harmonics $\Psi_{v\alpha L M}(\gamma,\Omega)$.
The matrix elements of physical operators
with respect to this basis, both for the Hamiltonian and for
electromagnetic transitions, can be computed directly from the
$\grpsochain$ Clebsch-Gordan
coefficients~\cite{rowe2004:spherical-harmonics,caprioXXXX:gammaharmonic,rowe-INPREP}.
Seniority quantum numbers $v\<\leq50$ amply suffice for convergence of
the calculations shown.  

The nature of the spectrum obtained depends both on the shape of the
potential (determined by $\xi$) and on the depth of the potential
(determined by $\chi$).    The
low-lying states form quasi-bands which may be roughly identified with
the $\gamma$ vibrational excitation ($K\<=2$) and two-phonon $\gamma$ excitations
($K\<=0$ and $4$).  For each
calculation in Fig.~\ref{fig-gammarotor-schemes}, $\chi$ is chosen to give
$E(2^+_\gamma)/E(2^+_1)\<\approx10$, appropriate to the well-deformed
rare earth nuclei.  Principal spectroscopic properties considered here
include the
band energies, the detailed level spacings within the bands, and the
interband electric quadrupole transition strengths.  

With the onset of triaxiality (increasing $\xi$), the two-phonon
energy anharmonicities evolve from slightly negative
($E_{\gamma\gamma}/E_{\gamma}\<<2$) for $\xi\<=0$
[Fig.~\ref{fig-gammarotor-schemes}(a)] to positive
($E_{\gamma\gamma}/E_{\gamma}\<>2$) [Fig.~\ref{fig-gammarotor-schemes}(c)].
The anharmonicity of the excited $K\<=0$ quasi-band rises more rapidly
than that of the $K\<=4$ quasi-band.  Qualitatively, this is
consistent with evolution towards a $\gamma$-stiff, adiabatic triaxial
rotor~\cite{rowe-INPREP,rowanwood-INPRESS}, for which the $K\<=4$ quasi-band is a
triaxial \textit{rotational} excitation and the $K\<=0$ quasi-band is a $\gamma$
\textit{vibrational} excitation.  
\begin{figure}
\begin{center}
\includegraphics*[width=0.7\hsize]{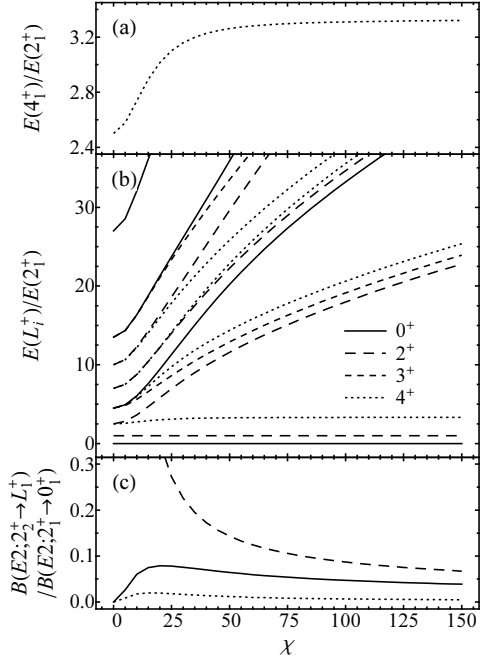}
\end{center}
\caption{Evolution of spectroscopic properties with $\gamma$ stiffness
for the angular Hamiltonian~(\ref{eqn-gammarotor}), for a purely axial
potential ($\xi\<=0$).  Quantities shown are (a)~the energy ratio
$E(4^+_1)/E(2^+_1)$, (b)~excitation energies of low-lying levels,
normalized to $E(2^+_1)$, and (c)~electric quadrupole transition strengths from the
quasi-$\gamma$ band head to the ground band members, normalized to
$B(E2;2^+_1\rightarrow0^+_1)$.  }
\label{fig-gammarotor-evoln}
\end{figure}

It is essential to observe that the stiffness of the potential around
its minimum in $\gamma$ determines not only the $\gamma$ vibrational
energy scale but also how well confined the wave function is with
respect to $\gamma$.  Thus, within the framework of the Bohr
Hamiltonian, the $\gamma$ band energy [more specifically, the energy
ratio $E(2^+_\gamma)/E(2^+_1)$, or separation of vibrational and
rotational energy scales] and the $\gamma$ softness of the wave
function are inextricably linked.  From
Fig.~\ref{fig-gammarotor-schemes}(a,inset), it is seen that for
$E(2^+_\gamma)/E(2^+_1)\<\approx10$ the
$\gamma$ confinement is weak, and that the range of energetically
accessible $\gamma$ values increases significantly for successive
phonon excitations.  Confinement is almost nonexistent at the energy
of the two-phonon excitations.

Consequently, dynamical $\gamma$ deformation plays a major role in the
calculated structure, as reflected in
significant deviations from ideal rotational behavior in the
spectroscopic predictions.
Level energies within the $\gamma$ quasi-band
[Fig.~\ref{fig-gammarotor-schemes}(a)] follow a gently $\gamma$-soft
staggering pattern [$2(34)(56)\ldots$].  (The relation between the
$\gamma$ excitation energy and residual level energy staggering was
noted for transitional nuclei in Ref.~\cite{caprio2005:axialsep}.)
The deviations are even more pronounced for the two-phonon
quasi-bands.  Note especially the near doubling of the rotational
constant, or rotational energy spacing scale, for the two-phonon bands
relative to the ground state band.

With increasing $\xi$, the level energies within
the $\gamma$ band progress from $\gamma$-soft staggering to the
pattern associated with triaxial rotation
[$(23)(45)\ldots$]~\cite{davydov1958:arm-intro}.  This may be seen
most clearly from plots of the level energy second difference
$S(L)\<\equiv\bigl[[E(L)-E(L-1)]-[E(L-1)-E(L-2)]\bigr]/E(2^+_1)$
[Fig.~\ref{fig-gammarotor-schemes}(d--f)], which has minima at even
$L$ for $\gamma$-soft staggering and at odd $L$ for triaxial
staggering.  As surveyed in
Ref.~\cite{mccutchan2007:gamma-staggering}, the data for most
rotational nuclei yield $S(L)$ plots which are either $\gamma$-soft
[Fig.~\ref{fig-gammarotor-schemes}(d)] or near-constant
[Fig.~\ref{fig-gammarotor-schemes}(e)].

The relation between $\gamma$ softness and spectroscopic properties is
more systematically and quantitatively examined in
Fig.~\ref{fig-gammarotor-evoln}.  For
Hamiltonian~(\ref{eqn-gammarotor}), at fixed $\xi$, the parameter
$\chi$ controls the depth and hence $\gamma$ stiffness of the
potential.  The evolution of energy and transition strength prediction
with respect to $\chi$ is shown in Fig.~\ref{fig-gammarotor-evoln} for
the pure $V(\gamma)\<\propto(1-\cos3\gamma)$ potential ($\xi\<=0$).
Note especially the correlation between the $\gamma$ band energy
[Fig.~\ref{fig-gammarotor-evoln}(b)] and the ground state band energy
ratio $E(4^+_1)/E(2^+_1)$ [Fig.~\ref{fig-gammarotor-evoln}(a)], which
varies from $2.5$ for $\gamma$-soft rotation to $3.33$ for rigid axial
rotation.  This ratio is commonly taken as an indicator of rotational
adiabaticity.  (The quantitative details are affected also by the
$\beta$ degree of freedom.)  The evolution of multi-phonon band
energies, approaching harmonicity for large $\chi$, can also be traced
in Fig.~\ref{fig-gammarotor-evoln}.  The electric quadrupole branching
ratios, shown for the $2^+_\gamma$ state in
Fig.~\ref{fig-gammarotor-evoln}(c), approach the Alaga rule
ratios~\cite{alaga1955:branching} of the adiabatic axial rotor, but
only slowly, as the $\gamma$ stiffness increases.
\begin{figure}
\begin{center}
\includegraphics*[width=\hsize]{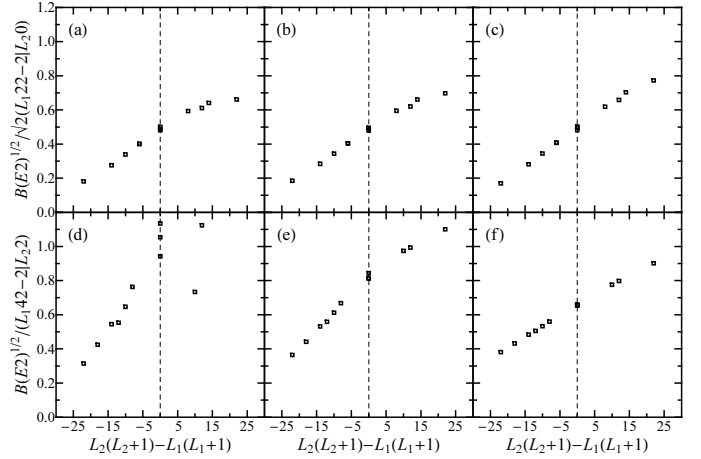}
\end{center}
\caption{
Interband transition amplitudes, from the $\gamma$ quasi-band to the
ground state band~(top) and from the $K\<=4$ $\gamma\gamma$ quasi-band
to the $\gamma$ quasi-band~(bottom), in Mikhailov form.  Values are
shown for the calculations of Fig.~\ref{fig-gammarotor-schemes}, with
$\xi\<=0$ ($\chi\<=50$)~(left), $\xi\<=0.5$ ($\chi\<=100$)~(middle),
and $\xi\<=0.8$ ($\chi\<=500$)~(right).  The values shown are for
transitions between levels with $L\<\leq6$, normalized to
$B(E2;2^+_1\rightarrow0^+_1)\<\equiv1$.
}
\label{fig-gammarotor-mikhailov}
\end{figure}

The phenomenological analysis of electromagnetic transition strengths
in rotational nuclei is founded upon the reduction of these strengths
to a single intrinsic electromagnetic matrix element and single mixing
parameter between each pair of bands, according to the Mikhailov
mixing formalism~\cite{mikhailov1966:mixing-APS}.  Within this
framework, all transition amplitudes fall on a straight line on an
appropriate (Mikhailov) plot, and the intrinsic matrix elements and
mixing parameter are identified from the slope and
intercept~\cite[Sec.~4-4]{bohr1998:v2}.  The electric quadrupole
transition strengths for the fully-converged Bohr Hamiltonian
calculations described above (Fig.~\ref{fig-gammarotor-schemes}) are
shown in Mikhailov form in Fig.~\ref{fig-gammarotor-mikhailov}.
Deviations from linearity are significant for transitions involving
the two-phonon excitations in the $\xi\<=0$ calculation
[Fig.~\ref{fig-gammarotor-mikhailov}(d)], as might be expected from
the substantial $\gamma$ softness already noted.  Otherwise the
transition amplitudes at least approximately follow an essentially
linear pattern.  It is therefore meaningful to extract effective
intrinsic matrix elements and mixing parameters for comparison with
experiment.  The matrix elements for the $\gamma\rightarrow g$
transitions and for the $\gamma\gamma\rightarrow
\gamma$ transitions (relative to $\gamma\rightarrow g$) are listed in
Table~\ref{tab-ime}.  The adiabatic harmonic
values~\cite{rowanwood-INPRESS} and a previous estimate [$\grpy{5}$] for the
``critical'' ($\xi\<=0.5$) case~\cite{iachello2003:y5} are shown for
comparison.
\begin{table}
\caption{Effective interband intrinsic matrix elements, as extracted from
the calculations of
Fig.~\ref{fig-gammarotor-mikhailov}.  The harmonic
axial~\cite{rowanwood-INPRESS} and $\grpy{5}$
triaxial~\cite{iachello2003:y5} estimates are included for comparison,
along with the experimental values for
$^{162}\mathrm{Dy}$~\cite{aprahamian2006:162dy-grid}.  }
\label{tab-ime}
\begin{tabular}{lr@{}lr@{}lr@{}l}
\hline  
\\[-8pt] 
 &
\multicolumn{2}{l}{$\frac{\me{0_g}{\calM}{2_\gamma}}{B(E2;2^+_1\rightarrow0^+_1)^{1/2}}$}
&
\multicolumn{2}{l}{$\frac{\me{2_\gamma}{\calM}{4_{\gamma\gamma}}}{\me{0_g}{\calM}{2_\gamma}}$}
&
\multicolumn{2}{l}{$\frac{\me{2_\gamma}{\calM}{0_{\gamma\gamma}}}{\me{0_g}{\calM}{2_\gamma}}$}
\\
\hline
$\xi\<=0$ ($\chi\<=50$)
& ~~~~~~0&.51 & ~~$\rtrim\sim$1&.9 & ~~$\rtrim\sim$0&.8\\
$\xi\<=0.5$ ($\chi\<=100$)
& 0&.52 & 1&.90 & 0&.54\\
$\xi\<=0.8$ ($\chi\<=500$)
& 0&.54 & 1&.46 & 0&.44\\
\hline
Harmonic & \multicolumn{2}{c}{---} & 1&.41 & 1&\\
$\grpy{5}$ & \multicolumn{2}{c}{---} & 1&.23 & 0&.73\\
\hline
\\[-8pt] 
$^{162}\mathrm{Dy}$ & 0&.241(3) & 0&.99(3) & 0&.54(3)$^a$
\\\hline  
\end{tabular}
\\
$^a$ For the $K^\pi\<=0^+$ excitation  at
$1400\,\mathrm{keV}$
excitation energy.
\end{table}


The nucleus $^{162}\mathrm{Dy}$ has recently been the subject of
detailed spectroscopic study by Aprahamian \textit{et
al.}~\cite{aprahamian2006:162dy-grid}, yielding extensive sets of
experimental values for interband electric transition strengths.  The
experimental intrinsic matrix elements are given in
Table~\ref{tab-ime}.  The $\gamma$ band and candidate two-phonon
$\gamma$ band energies for $^{162}\mathrm{Dy}$ closely match those of
the $\xi\<=0$ ($\chi\<=50$) calculation of
Fig.~\ref{fig-gammarotor-schemes}(a).  The measured $\gamma\rightarrow
g$ strength is about a factor of two lower than the Bohr Hamiltonian
predictions, and the $K\<=4$  $\gamma\gamma\rightarrow
\gamma$ strength is yet a factor of two lower again, \textit{i.e.}, even taken relative to this already reduced
$\gamma\rightarrow g$ strength.  Moreover, the observed level energies
in $^{162}\mathrm{Dy}$~\cite{aprahamian2006:162dy-grid} conform much
more closely to adiabatic rotational $L(L+1)$ spacings than expected
from the Bohr Hamiltonian calculation
[Fig.~\ref{fig-gammarotor-schemes}(a)].  Although we do not include a
detailed quantitative analysis here, for the $\gamma$ band staggering,
compare Fig.~3(d) of Ref.~\cite{mccutchan2007:gamma-staggering} with
the present Fig.~\ref{fig-gammarotor-schemes}(d).  The
excited bands in $^{162}\mathrm{Dy}$ are actually observed to exhibit
\textit{decreased} rotational constants, relative to the ground state
band, in stark contrast to the present calculation (a similar
discrepancy is noted~\cite{warner1981:168er-ibm} in comparison of interacting boson model~\cite{iachello1987:ibm}
predictions with data).
A more detailed comparison requires exploration of the
interaction of the $\beta$ and $\gamma$ degrees of freedom.
\begin{figure}
\begin{center}
\includegraphics*[width=\hsize]{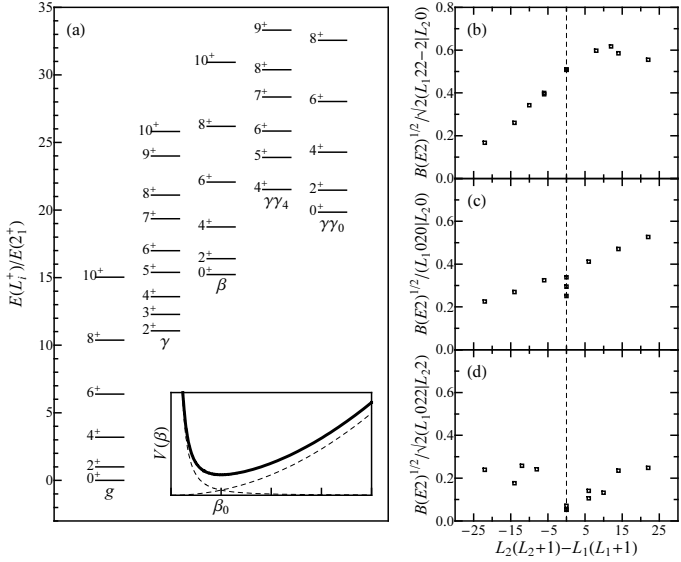}
\end{center}
\caption{
Exact numerical diagonalization of the Bohr Hamiltonian for a
well-deformed rotor with the full dynamics of the $\beta$, $\gamma$,
and rotational degrees of freedom.  Confinement in $\beta$ is provided
by a Davidson potential ($\beta_0\<=3$) with $\gamma$ stiffness
parameter $\chi\<=5$ (see text).  (Left)~Level energies, normalized to
$E(2^+_1)$, with an illustration of the Davidson potential
(inset).  The dashed lines indicate the asymptotic forms
$\beta^{-2}$ and $\beta^{2}$ of the potential, at small and large
$\beta$, respectively.
(Right)~Transition amplitudes: (b)~$\gamma\rightarrow g$,
(c)~$\beta\rightarrow g$, and (d)~$\beta\rightarrow\gamma$.  The
values shown are for transitions between levels with $L\<\leq6$,
normalized to $B(E2;2^+_1\rightarrow0^+_1)\<\equiv1$.  }
\label{fig-davidson}
\end{figure}

Let us briefly examine the results of a  calculation involving the
full dynamics of the Bohr Hamiltonian~(\ref{eqn-bohr}), including the
$\beta$ degree of freedom.  An essentially unlimited variety of
combined $\beta$ and $\gamma$ dependences (\textit{e.g.}, $\beta^m
\cos^n 3\gamma$) may be used in the ACM
potential~\cite{rowe-INPREP}.  For illustration, we consider a
Davidson potential for $\beta$ [Fig.~\ref{fig-davidson}(a,inset)],
together with the simplest axial $\gamma$ confinement from above, so
$V(\beta,\gamma)\propto(\beta_0/\beta-\beta/\beta_0)^2+\chi(1-\cos3\gamma)$.
Without loss of generality, we may take the proportionality constant and the
inertial constant $\hbar^2/(2B)$ to be unity, if only energy and $E2$ strength ratios
are to be considered~\cite{caprio2003:gcm}.  This leaves freedom only in the $\beta$ stiffness ($\beta_0$) and $\gamma$
stiffness ($\chi$).  A fully converged ACM calculation, for
$\beta_0\<=3$ and $\chi\<=5$, is shown in Fig.~\ref{fig-davidson}(a),
with quasi-band assignments indicated.  Fewer than five
$\grpsu{1,1}$ basis functions are required for convergence.
Electric quadrupole transitions connecting the ground, $\beta$, and
$\gamma$ bands are shown in Fig.~\ref{fig-davidson}(b--d).  The zero
in the Mikhailov plot for the $\beta\rightarrow\gamma$ transitions
[Fig.~\ref{fig-davidson}(d)] reflects the zero expected [at
$L_2(L_2+1)-L_1(L_1+1)\<=4$] if simple phonon selection rules hold on
the intrinsic matrix elements (see Fig.~12 of Ref.~\cite{casten1983:ibm-deformed}).

The possibility of exact diagonalization of the Bohr Hamiltonian for
essentially arbitrary $\beta$ and $\gamma$ stiffness, by means of the
algebraic collective model, opens the door for direct comparison of
the Bohr Hamiltonian predictions with experiment throughout the range
of possible dynamics for the nuclear quadrupole degree of freedom.  At
a phenomenological level, this permits meaningful tests of the Bohr
Hamiltonian for general rotor-vibrator nuclei.  More fundamentally, in
the context of nuclear many-body theory, it is essential to know the
limitations and necessary modifications to the Bohr framework, since
microscopic descriptions of nuclear collectivity rely upon a reduction
of the many-body problem to one involving effective collective degrees
of freedom, typically those of the Bohr
description~\cite{eisenberg1976:v3,zelevinsky2002:soft-microscopic-icns02}
or its symplectic
generalization~\cite{rowe1985:micro-collective-sp3r}.

Notably, these preliminary results suggest that the Bohr description
quantitatively overpredicts and even qualitatively mis-predicts the
nature of deviations from adiabatic rotational structure, and that
these deviations result from dynamical $\gamma$ softness inherently
incorporated into the description.  The Bohr description also strongly
overpredicts $\gamma$ phonon interband transition strengths (although
this could already be anticipated from, \textit{e.g.},
rotation-vibration model estimates).  A basic question is therefore
the extent to which the discrepancies lie in the details
of the description (\textit{e.g.}, restriction of the kinetic energy
operator to quadratic order in the momenta) or in a fundamental
failure of the collective quadrupole degree of freedom to account for
the observed phenomena.


Valuable discussions with N.~V.~Zamfir, D.~J.~Rowe, F.~Iachello,
S.~Frauendorf, and A.~Aprahamian are gratefully acknowledged.  This
work was supported by the US DOE under grant DE-FG02-95ER-40934.

\vfil


\providecommand{\ELSEVIER}{}
\ELSEVIER\newcommand{\identity}[1]{{#1}}


\end{document}